\documentclass[%final            % use final for the camera ready runs
  ,draft]{aipproc}
\usepackage{amsmath}
\usepackage{amssymb}
\usepackage[english]{babel}

\layoutstyle{8x11double}

\begin{document}

\title{Nucleon Emision Off Nuclei Induced By Neutrino Interactions}


\author{M. Valverde}{
  address={Research Center for Nuclear Physics (RCNP), Osaka University, Ibaraki 567-0047, Japan}
}

\author{J. Nieves}{ address={Instituto de F\'\i sica Corpuscular
    (IFIC), Centro Mixto CSIC-Universidad de Valencia, Institutos de
    Investigaci\'on de Paterna, Aptd. 22085, E-46071 Valencia, Spain}}

\author{J. E. Amaro}{ address={Dpto. de F\'\i sica At\'omica,
    Molecular y Nuclear, Universidad de Granada,Facultad de Ciencias,
    Campus Fuentenueva S/N, E-18071 Granada, Spain}}

\author{M. J. Vicente-Vacas}{ address={Dpto. de F\'\i sica Te\'orica e Instituto de F\'\i sica Corpuscular
    (IFIC), Centro Mixto CSIC-Universidad de Valencia, Institutos de
    Investigaci\'on de Paterna, Aptd. 22085, E-46071 Valencia, Spain}}

\begin{abstract}
We make a review of the main nuclear effects that affect
neutrino-nucleus cross sections.  We discuss how the different models
in the literature try to describe these different effects, and thus
try to compare between them. We focus on the quasi-elastic reaction in
the neutrino energy region of around 1 GeV, where recent data from
MiniBoone are available. Among the issues discussed are the different
treatment of medium corrections to initial and final state nucleon
wave functions and the problem of the rescattering of ejected
nucleons.
\end{abstract}

\maketitle

%%%%%%%%%%%%%%%%%%%%%%%%%%%%%%%%%%%%%%%%%%%%
%% MAINMATTER
%%%%%%%%%%%%%%%%%%%%%%%%%%%%%%%%%%%%%%%%%%%%

\section{Introduction}

The study of neutrinos is playing a very relevant role in current
research in nuclear, astro and particle physics. One of these major
topics is neutrino oscillations, that since its discovery 10 years ago
by the Super Kamiokande collaboration \cite{Kamiokande,kamiokande2}
have evolved and is now reaching the realm of precision experiments
\cite{Hagiwara:2005pe}. This new generation of precision experiments
is no longer hindered by statistical error, but is dominated by
systematic uncertainties, one of the most important of these
systematic errors being the neutrino-nucleus cross section. However
neutrinos are a neutral, weakly interacting particle, thus its
detection must rely on the observation of secondary particles that
appear after the scattering of the incoming neutrino with one of the
nuclei of the passive part of the detector set up.  The study of
neutrino oscillation physics requires a good determination of the
incoming neutrino energy. However accelerator neutrino beams are
produced by the muon decay, thus being far from monochromatic. The
determination of the energy of a detected neutrino and the nature of
the collision thus can be only done through kinematic reconstruction
from the produced particles.

A proper knowledge of the neutrino-nucleus cross sections is therefore
required for the experiment analysis. Nevertheless most of the codes
and models used for the modelling of these processes in the analysis
of experimental data are based on Fermi Gas models, namely the famous
Marteau model \cite{Marteau:1999kt}.  However most of these models are
known from the experience on electron scattering physics to fail to
describe the existing experimental data.

These reasons have motivated the interest of the theoretical nuclear
physics community on the subject of neutrino-nucleus scattering. There
is a general consensus among the community that a simple Fermi Gas
model, widely used in the analysis of neutrino oscillation
experiments, is no longer good enough for the level of precision
required for neutrino experiments. Thus many different approaches has
been proposed to model these reactions. In this talk I will describe
the main nuclear physics effects that, based on the experience from
electron scattering physics, are expected to arise. We will also try
to review a few of the proposed nuclear models. For the sake of
simplicity we will focus on the quasi-elastic reactions at neutrino
energies from a few hundred MeV to around 1 GeV which are of interest
to MiniBoone and T2K. Of course at these experiments pion production
processes play an important role, however the pion physics involved in
these processes is somewhat beyond the scope of this contribution. For
a more general look at this neutrino interactions issues please look
at the contribution of L. Alvarez Ruso in this same proceedings
\cite{LuisARuso}.

\section{Nuclear effects in inclusive processes}

In this section we introduce a model for the quasi-elastic inclusive charged current process 
\begin{equation}
\nu_l + \, ^Z\! A \to l^- + X
\end{equation}
where the only detected particle being the outgoing lepton and
therefore one must sum over all possible final hadron states, here
denoted by $X$.  In the case of neutral current processes the outgoing
lepton part is a neutrino. So in order to get information about the
process some of the hadronic products must be detected, be it a
nucleon (usually proton) or pion. The residual nucleus is not detected
in the Cerenkov radiation experiments so a sum over nuclear states is
also needed, what is usually called semi-inclusive observables. In
this talk we will focus in the situation where the only detected
hadron is a nucleon. If we were to include also pions we should take
into account the production of pions in the primary $\nu$-nucleon
vertex.  We will follow the model of references \cite{Nieves:2004wx}
and \cite{Nieves:2005rq} which start from a local density Fermi Gas
model, but in top of it they include a whole lot of nuclear
effects. We will refer as this model as the Valencia-Granada model.
Actually this model is actually based on a previous work dealing with
electron scattering \cite{Gil:1997bm} that was able to reasonably
describe the available experimental data.

\begin{figure}
  \includegraphics[width=.4\textwidth]{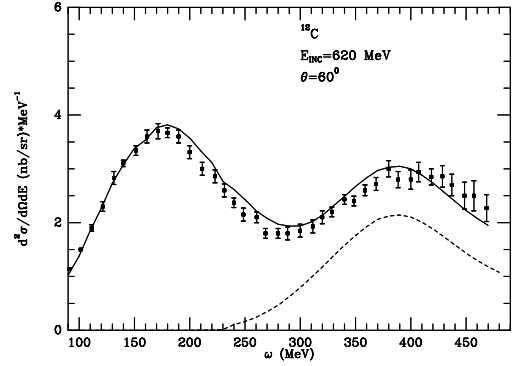}
  \caption{Double differential cross section for the inclusive process
    $e^- + C^{12}\to e^- + X$. Picture taken from \cite{Gil:1997bm}}
  \label{fig:gil}
\end{figure}

Due to the difficult nature of neutrino cross sections experiments
(that sometimes rely on some neutrino interaction model), it is very
important to validate any neutrino-nucleus interaction model against
electron-nucleus scattering experiments. As can be seen in Fig.\ref{fig:gil}
the electron version of the Valencia-Granada model describes rather
well the existing experimental data. In this figure we can observe the
main features of this kind of reactions. A large broad peak at low
transfered energy $\omega$ (the quasi-elastic peak), and a second lower peak
that is associated with the $\Delta(1232)$ production.
A major feature of this model is that it is able to correctly describe
the region between the two major peaks.  This {\em gap} region is
underestimated by most models, as they usually do not take into
account any process beyond the dominant absorption by one nucleon
(quasi-elastic) and delta production. However one must notice the
existence of additional processes like non-resonant pion production
and boson absorption by two nucleons. Thus in order to properly
describe inclusive processes it is clearly needed the inclusion of
additional non-resonant mechanisms. The first step towards the
inclusion this processes in the framework of a model of
neutrino-nucleus scattering is having an adequate model for
neutrino-nucleon scattering in free space (that is, with no nuclear
medium effects).  Many interesting approaches are being developed to
tackle this problem, see {\em e.g.}  Ref.\cite{Hernandez:2007qq}.

The differential cross section for the neutrino collision can be written
\begin{equation}
\frac{d^2\sigma_{\nu l}}{d\Omega(\hat{k^\prime})dE^\prime_l} =
\frac{\lvert\vec{k}^\prime\rvert}{\lvert\vec{k}\rvert}\frac{G^2}{4\pi^2} 
L_{\mu\sigma}W^{\mu\sigma} \label{eq:sec}
\end{equation}
with $L$ and $W$ the leptonic and hadronic tensors, respectively. The
leptonic tensor is well known and is obtained from the weak
interaction in the Fermi contact approximation.  On the other hand,
the inclusive CC nuclear cross section is related to the imaginary
part of the neutrino self-energy in the medium by:
\begin{equation}
\sigma = - \frac{1}{\lvert\vec{k}\rvert} \int\,d^3\vec{r} \; {\rm Im}\Sigma_\nu (k;\rho(r)) 
\end{equation}
We obtain the imaginary part of the neutrino self-energy in the medium
${\rm Im}\Sigma_\nu$ by means of the Cutkosky's rules. We obtain for
$k^0 > 0$
\begin{equation}
{\rm Im} \Sigma_\nu(k) = \frac{8G}{\sqrt 2 M^2_W}\int \frac{d^3
  k^\prime}{(2\pi)^3 }\frac{\Theta(q^0) }{2E^{\prime}_l} 
~ {\rm Im}\left\{ \Pi^{\mu\nu}_W(q;\rho) L_{\nu\mu} \right\} 
\label{eq:ims}
\end{equation}
and thus, the hadronic tensor is basically an integral over the
nuclear volume of the $W$-boson self-energy
$\Pi_W^{\mu\nu}\left(q\,;\rho\right)$ inside the nuclear medium. In
general we can then take into account the different in-medium effects
and reaction mechanism modes ($W$ absorption by one nucleon or by a
pair of nucleons, pion production, resonance excitation\ldots) by
including the correspondent diagrams in the $W$ self-energy diagram.
We will focus in the charged current quasi-elastic process, that
corresponds to the $W$ absorption by one nucleon.  In general we
obtain that the hadron tensor can be expressed (up to some constant)
as:
\begin{multline}
W^{\mu\nu} = \int d^3\vec{r}\int \frac{d^3\vec{p}}{(2\pi)^3}\int_{\mu-q^0}^{\mu} d\omega \,
A^{\nu\mu}(p,q)|_{p^0=\bar{E}(\vec{p})} \\
S_h\left(\omega,\vec{p};\rho\right) S_p\left(q^0+\omega,\vec{p}+\vec{q};\rho\right) \, .
\end{multline}
In this expression the tensor $A^{\nu\mu}$ contains all the
information related to the neutrino-nucleon interaction. The $S_p$ and
$S_h$ are the particle and hole nucleon spectral functions and contain
the information on the nucleon wave functions in the final and initial
nuclear state respectively. $\mu$ is the chemical potential. In a
simple local density Fermi Gas the nucleons are on mass shell and thus the hole spectral functions take the very simple form:
\begin{equation}
S_h\left(\omega,\vec{p};\rho\right) = \delta(\omega - E(\vec{p}))\Omega(E_F - E(\vec{p}))
\end{equation}
and an analogue expression for the particle one.  This expression leads to a
description that is completely equivalent to the usual Fermi gas model
used in the literature. The only nuclear physics effects that are
taken into account are the Pauli blocking effect (see the $\Theta$
function) and the Fermi motion of the nucleons, whose momenta are
approximated to be distributed uniformly.

\begin{figure}
  \includegraphics[width=.4\textwidth]{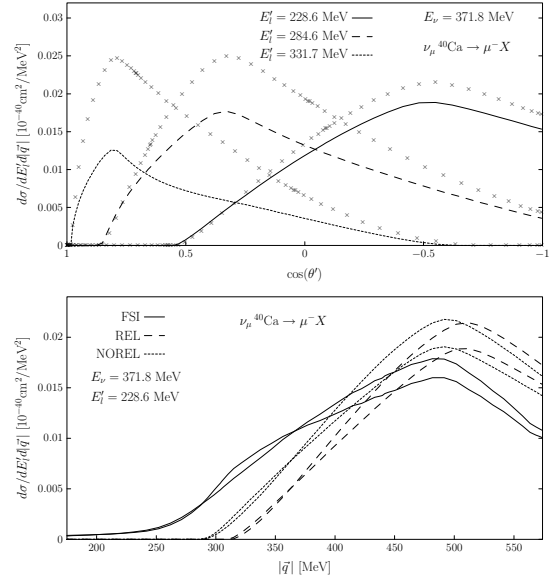}
  \caption{$\nu_e$ and $\bar{\nu}_e$ inclusive quasi-elastic cross
    sections in oxygen as a function of the transferred energy, at two
    values of the transferred momentum. We show results for
    relativistic (REL) and non-relativistic nucleon kinematics. In
    this latter case, we present results with (FSI) and without
    (NOREL) FSI effects. For the three cases, we also show the effect
    of taking into account RPA correlations (lower lines at the
    peak). See Ref.~\cite{Nieves:2004wx} for further details.}
  \label{fig:fsi}
\end{figure}

However, this is well known to be an oversimplificated model for the
electron scattering process, and we expect it to be so also for the
neutrino process. Thus we improve our model by including realistic
spectral functions,
\begin{multline}
S_{p,h}(\omega,\vec{p}\,;\rho) = \\
\mp\frac{1}{\pi}\frac{{\rm Im}\Sigma(\omega,\vec{p}\,;\rho)}
 {\left[\omega-{\bar E}(\vec{p}\,)-{\rm Re}\Sigma(\omega,\vec{p}\,;\rho)  \right]^2 + \left[{\rm Im}\Sigma(\omega,\vec{p}\,;\rho)\right]^2}
\end{multline}
with $\omega\ge \mu$ or $\omega\le \mu$  for $S_p$ and $S_h$,
respectively.  The chemical potential $\mu$ is determined by
\begin{equation}
\mu = M + \frac{k_F^2}{2M} + {\rm Re}\Sigma(\mu, k_F)
\end{equation}
where in Valencia-Granada model the reference~\cite{FO92} was followed
for the nucleon self-energy $\Sigma(\mu,k_F)$. Notice that both
particle and hole self-energies are included in this approach, in
contrast with other models in the literature that only include the
effect of nucleon wave functions in the hole states. It is also very
important to notice how in the limit $\Sigma \to 0$ we recover the
expressions of a non-interacting Fermi Gas model. This is an obvious
point that should be tested in all models for lepton scattering off
nuclei. The effect of particle (that is final state nucleons) spectral
functions is often defined in the literature as final state
interactions (FSI). This effect (see Fig.\ref{fig:fsi}) usually produces a
broadening of the nuclear response and a reduction of the response at
the peak, however the total response is not much affected, specially
when RPA corrections (see next paragraph) are also taken into account.

\begin{figure}
  \includegraphics[width=.4\textwidth]{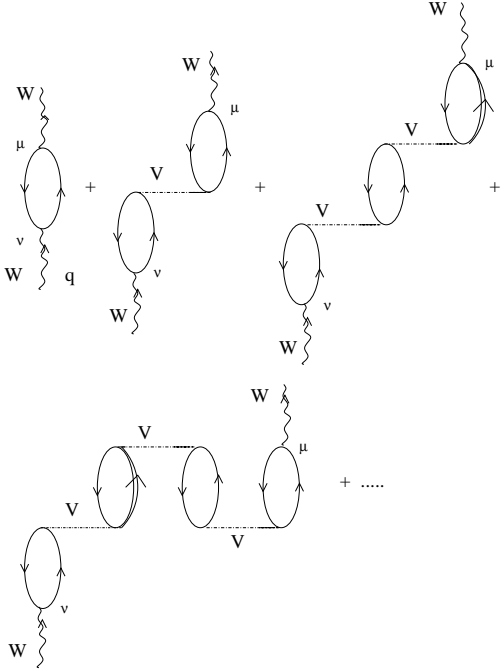}
  \caption{Set of irreducible diagrams responsible for the
    polarization (RPA) effects in the $1p1h$ contribution to the $W$
    self-energy.}
  \label{fig:rpa}
\end{figure}

Furthermore the excited nuclear states are expected to be correlated
by means of the nucleon-nucleon interaction.  We model this effect by
including a series of particle-hole excitations, see
Fig.\ref{fig:rpa}, of the RPA type. The inclusion of this diagrams
modify the expression for the tensor $A^{\nu\mu}$ and induces a
reduction of the cross section, specially at low $Q^2$ kinematics.  We
use an effective Landau-Migdal $ph-ph$ interaction where in the
vector-isovector channel
($\vec{\sigma}\cdot\vec{\sigma}\vec{\tau}\cdot\vec{\tau}$ operator) we
use an interaction with explicit $\pi$ meson (longitudinal) and $\rho$
meson (transverse) exchanges that also includes $\Delta(1232)$
degrees of freedom.

This point has been applied in \cite{AlvarezRuso:2009ad} to the
MiniBoone experiment \cite{Katori:2009du} following the prescriptions
of our model. In this reference it was found that the inclusion of
this RPA correlations improves the description of the cross section
measurements in the MiniBoone experiment, without including unphysical
parameters, like effective Fermi momentum\ldots
\begin{figure}
  \includegraphics[width=.4\textwidth]{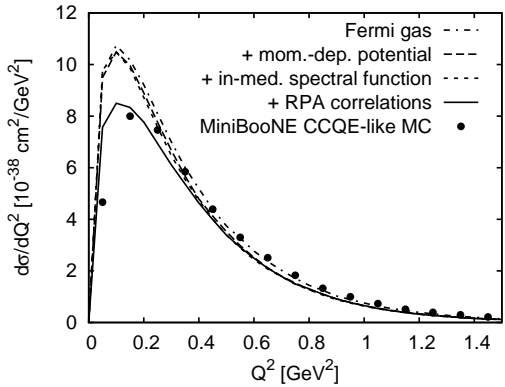}
  \caption{Taken from Ref.~\cite{Leitner:2008fg}.}
  \label{fig:leitner}
\end{figure}

\section{Semi-inclusive observables: Nucleon rescattering}

In the previous section we have focused in processes where the only
detected particle is the outgoing (charged) lepton. However sometimes
more information from the process is obtained from hadrons. Actually
these are the only possible particles to be detected in neutral current
processes. In this process a new effect must be taken into account in
top of the ones described in the previous section. This is the
rescattering of outgoing hadrons in its way out of the
nucleus. Actually in the previous model the nucleon that interacts
with the neutrino is put on mass shell and goes out of the
nucleus. However it is well known that this nucleon strongly interacts
with the other nucleons in the medium and can be deflected,
inducing the emission of secondary nucleons or, given enough energy,
pions. Of course this new processes do not change the total inclusive
cross section as described in the previous section. However in the
case of neutral currents it is necessary to properly describe this
processes as secondary particle emission processes can result in
background events, {\it e.g.} $\pi^0$ decay photons can mimic Cerenkov
radiation from electrons, or in other processes energy can be
transfered to undetected neutrons thus introducing problems in the
incoming neutrino kinematics reconstruction. For that reason it is
very important to properly model this rescattering processes.
A few different approaches have been proposed to describe this:
\begin{enumerate}
\item Distorted wave impulse approximation. In this models the
  outgoing nucleon wave function is calculated using a wave equation
  complex potential. The imaginary part of this potential removes all
  the events where the outgoing nucleon collides. This approach is
  fully quantum mechanical, however it has the major disadvantage that
  is only suitable to deal with fully exclusive observables where the
  final nuclear state is also observed.  These is because the optical
  potential is not unitary and thus it does not shuffle events from
  one channel to another (as should be done when dealing with
  semi-inclusive observables) but just remove those events where the
  nucleon undergoes a collision,changing its kinematics.  For that
  reason it is well known that this approach underestimates cross
  sections in semi-inclusive reactions.
\item Monte Carlo cascade models. This is the usual approach in which
  the trajectory of the ejected hadrons is simulated via a
  semi-classical Monte Carlo algorithm that takes into account changes
  of energy and momentum of the emitted nucleon, as well as the
  possibility of having secondary hadrons.
\item Transport model. Recently a new approach has been proposed by
  the Giessen group \cite{Leitner:2006ww}. In this approach the semi-classical
  transport equations are explicitly solved for all ejected hadrons,
  thus allowing for a rigorous tracking of all particles.
\end{enumerate}

In the following we shall focus on the model by the Valencia-Granada
group \cite{Nieves:2005rq}, which is a Monte Carlo cascade like
model. We shall use a simplified version of the model where pion
production processes are not included. In this model for any given
leptonic kinematic $q^\mu$, a point ($\vec{r}$) in the nucleus is
randomly selected where the gauge boson absorption takes place
according to the profile $d^5\sigma / d\Omega'dE'd^3\vec{r}$.  Then a
nucleon with a random momentum is picked up from the Fermi sea with a
given momentum $\vec{p}$. Its kinematics is determined via energy
conservation
\begin{equation}
  E = q^0 + \sqrt{\vec{p}^2 + M^2}
\end{equation}
and Pauli blocking effects are explicitly included.  The nucleon is
assumed to be in an average nucleon potential $V(r) =
k_F^2(\vec{r})/2M$ and then it is moved in discrete steps until it
leaves the nucleus. At each of these steps the possibility of
producing a secondary nucleon is explicitly taken into account by
means of the cross section
\begin{equation}
  \hat{\sigma}^{N_{1}N_{2}} = \int d\Omega_{CM}
  \frac{d\sigma^{N_{1}N_{2}}}{d\Omega_{CM}}C_T(q,\rho)
  \Theta\left(\kappa-\frac{\lvert\vec{p}\cdot\vec{p}_{CM}\rvert} {\lvert\vec{p}\rvert\lvert\vec{p}_{CM}\rvert}\right)
\end{equation}
where in-medium renormalization of the nucleon-nucleon interaction
($C_T(q,\rho)$) and Pauli blocking effects $\Theta\left(\kappa-
\lvert\vec{p}\cdot\vec{p}_{CM}\rvert
\ \lvert\vec{p}\rvert\lvert\vec{p}_{CM}\rvert\right)$ are explicitly
included.

\begin{figure}
\includegraphics[width=0.45\textwidth]{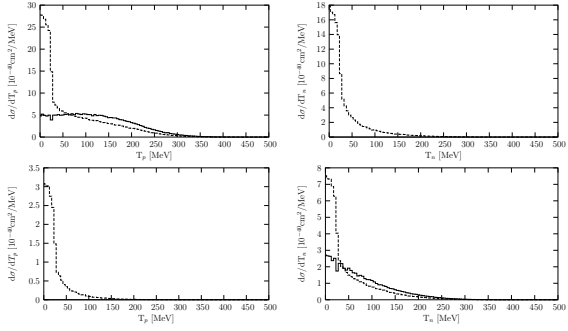}
\caption{\footnotesize Charged current $^{40}Ar(\nu,\mu^-+N)$ (upper
  panels) and $^{40}Ar(\bar{\nu},\mu^++N)$ (lower panels) cross
  sections as a function of the kinetic energy of the final
  nucleon. Left and right panels correspond to the emission of protons
  and neutrons respectively. The solid histogram shows results without
  FSI and the dashed one the full model. Please look
  Ref.~\cite{Nieves:2005rq} for further details.}
\label{fig:mcharged}
\end{figure}

The effect of this cascade algorithm in the spectra of outgoing nucleons can be easily appreciated
in Fig.\ref{fig:mcharged}.
The nucleons spectra produced by CC processes induced by muon neutrinos 
are shown in Fig.~\ref{fig:mcharged} for Argon. Of course neutrinos  
only interact via CC with neutrons and would emit protons, but these primary 
protons interact strongly with the medium and collide with other nucleons which
are also ejected. As a consequence there is  a reduction of the flux of high 
energy protons but a large number of secondary nucleons, many of them neutrons,
of lower energies appear.

%%%%%%%%%%%%%%%%%%%%%%%%%%%%%%%%%%%%%%%%%%%%%%%%
%% BACKMATTER
%%%%%%%%%%%%%%%%%%%%%%%%%%%%%%%%%%%%%%%%%%%%%%%%

\begin{theacknowledgments}
This research was supported by spanish DGI and european FEDER funds,
under contracts FIS2005-01143/FIS, FIS2006-3048, FPA2007-65748,
CDS2007-00042, by Junta de Castilla y Le\'on (Spain) under contracts
SA016A07 and GR12 and by the EU HadronPhysics2 project. M. Valverde
wishes to acknowledge a fellowship from the Japanese Society for the
Promotion of Science. 
%We also gratefully acknowledge permission from S. Boyd, S. Ditman and
%J. Sobczyk for using some pictures.

\end{theacknowledgments}

%% Please use U.S. style formatting if BibTeX is not used.
%%
%% For your convenience a BibTeX coded example is appended

\end{document}